\documentclass[prl,twocolumn,showpacs]{revtex4}
\usepackage{epsfig}
\usepackage{amsbsy}
\usepackage{amsmath}
\usepackage{graphicx}
\usepackage{latexsym}

\begin{document}

\title{Mode entanglement of electrons in the one-dimensional Frenkel-Kontorova model}
\author{Xiaoguang Wang$^{(1,2)}$, Haibin Li$^{(1,3)}$, and Bambi Hu$^{(1,4)}$}
\affiliation{1. Department of Physics and Center for Nonlinear Studies, Hong Kong Baptist University, Hong Kong, China.}
\affiliation{2. Department of Physics and Australian Centre of Excellence for Quantum
Computer Technology, \\
Macquarie University, Sydney, New South Wales 2109, Australia.}
\affiliation{3. Zhejiang Institute of Modern Physics, Zhejiang University, Hangzhou 310027, China.}
\affiliation{4. Department of Physics, University of Houston, Houston, Texas 77204-5005, USA.}

\date{\today}
\begin{abstract}
We study the mode entanglement in the one-dimensional Frenkel-Kontorova model, and found that behaviors
quantum entanglement are distinct before and after the transition by breaking of analyticity. 
We show that the more extended the electron is, the more entangled the corresponding state. Finally, 
a quantitative relation is given between the average square of the concurrence quantifying the degree 
of entanglement and the participation ratio characterizing the degree of localization.
\end{abstract}
\pacs{73.21.-b, 05.45.Mt, 03.65.Ud}
\maketitle

Quantum information science have emerged as an active interdisciplinary area between 
quantum mechanics and information theory~\cite{Nie00}. Recently, it was suggested that the 
quantum information science may offer an powerful approach to the study of nonlinear complex 
quantum systems~\cite{Aha99,Nie02,DiV00}. Specifically, there may be close connections between 
quantum entanglement theory and many-body theory~\cite{Aha99,Nie00}. These ideas motivate us to 
investigate nonlinear complex quantum systems by entanglement theory.

In this letter, we study quantum entanglement of electrons in 
the Frenkel-Kontorova (FK) model~\cite{FK}, a paradigm in nonlinear science, and address the 
effects of the transition by breaking of analyticity (one striking feature of the FK model) 
on behaviors of entanglement. 
The FK model has been used to model various kinds of physical systems such as 
an electron in a quasi-1D metal below the Peierls transition~\cite{FKK}.
The FK model describes a one-dimensional chain of atoms with harmonic nearest neighbor interaction 
placed in a periodic potential. Due to the competition between the two length scales, the spring length 
and the period of the on-site potential, the FK model
exhibits a rich complex phenomena~\cite{Braun,hl98}. 
It is shown by Aubry~\cite{au83} that there exist two different ground-state
configurations for an incommensurate chain, and the transition from
one configuration to another is driven by a single parameter $K$. These two incommensurate
configurations correspond to invariance circle and cantorus of the
standard map~\cite{ch79}, respectively. 

The electronic properties such as the energy spectrum and quantum diffusion in the FK model have been studied~\cite{Tong02}. 
Quite recently, entanglement properties
were studied in the Harper model~\cite{Harper}, another paradigm of nonlinear science, 
and some connections are revealed between entanglement and localization~\cite{Lak03}.  
In contrast to the Harper model that have
been often used to study electron properties in incommensurate systems~\cite{Sokoloff,Harper},  the FK model has two control parameters $K$ and $\lambda$, which leads to more rich physics. Moreover, the Harper model exhibits a symmetry of self-duality, whereas 
the FK model does not. Next, we study ground-state entanglement properties of the FK model, and find that 
the entanglement changes drastically when one goes from one configuration of atoms to another. 

Let us start by recalling some basic facts about (spinless) fermions on a
lattice of atoms. Consider $N$ local fermionic modes (LFMs) -- sites which can be either empty
or occupied by an electron~\cite{Bravyi}. In the second-quantized picture the basic objects are the creation and annihilation operators $c_n^{\dagger }$ and $c_n$ of $n$-th LFM,
satisfying the canonical anti-commutation relations
\begin{equation}
[c_n,c_m]_{+}=0,\quad[c_n,c_m^{\dagger }]_{+}=\delta _{nm}.  \label{car}
\end{equation}
The Hilbert space naturally associated to the $N$ LFMs, known as Fock space $%
{\cal H}_F$, is spanned by $2^L$ basis vectors $|n_1,...,n_{L}\rangle:=%
\prod_{l=1}^{N} (c_l^\dagger)^{n_l}\vert 0\rangle\,(n_l=0,1\,\forall\, l).$

{}From the above occupation-number basis it should be evident that ${\cal H}%
_F$ is isomorphic to the $N$--qubit space. This is easily seen by defining
the mapping~\cite{Fermion} 
\begin{equation}
\Lambda :=\prod_{l=1}^N(c_l^{\dagger })^{n_l}|0\rangle \mapsto \otimes
_{l=1}^N|n_l\rangle =\otimes _{l=1}^N(\sigma _l^{+})^{n_l}|0\rangle ,
\label{mappp}
\end{equation}
where $\sigma _l^{+}$ is the raising operator of $l$-th qubit.
This is a Hilbert-space isomorphism between ${\cal H}_F$ and ${\rm \kern%
.24em\vrule width.04emheight1.46exdepth-.07ex\kern-.30emC}^{\otimes \,L}$.
After this identification one can discuss entanglement of fermions by
studying entanglement of qubits. Clearly, this entanglement is relative to 
the mapping (\ref{mappp}). By defining new fermionic modes via automorphisms of the algebra (\ref{car}) 
one gives rise to different mappings between ${\cal H}_F$ and ${\rm \kern.24em\vrule width.04emheight1.46exdepth-.07ex\kern-.30emC}%
^{\otimes \,L}$ with an associated different entanglement. This simple fact
is one of the manifestations of the relativity of entanglement \cite{Paolo2}.

Consider an electron hopping in a 1D FK chain described by the following Hamiltonian
\begin{equation}
H=-t\sum_{n=1}^N (c_n^\dagger c_{n+1}+c_{n+1}^\dagger c_n) +\sum_{n=1}^N V_n c_n^\dagger c_n, \label{hamiltonian}
\end{equation}
where $t$ is a nearest-neighbor hopping integral which is set to 1 throughout the paper, and 
$
V_n=\lambda\cos(2\pi\sigma x_n^0)
$ 
is the on-site potential, which is  
controlled by the parameter $\lambda$ and the configuration $\{x_n^0\}$.  Here, $\lambda$ is the amplitude of the on-site potential, $\sigma=F_n/F_{n-1}$ is the inverse distance between two consecutive atoms for $K=0$, and  $\{F_n\}$ is a Fibonacci sequence and the series of truncated fraction $F_{n-1}/F_{n}$ converse to the inverse golden mean $(\sqrt{5}-1)/2$. 
The number of lattice sites is chosen to be $N=F_n$. 

The configuration $\{ x_n^0\}$ is the FK model is determined by minimizing the functional
\begin{equation}
U=\sum_n
\frac{1}{2}(x_{n+1}-x_n)^2+K[1-\cos(2\pi x_n)],
\label{fkp}
\end{equation}
where $K$ is a coupling constant controlling configurations of atoms. Moreover, the 
periodic boundary condition is assumed for Hamiltonian (\ref{hamiltonian}).

A general state of a single electron moving on the 1D FK chain is given by
\begin{equation}
|\Psi\rangle=\sum_{n=1}^N \psi_n |n\rangle=\sum_{n=1}^N\psi_n c_n^\dagger |0\rangle, \label{state}
\end{equation}
where $|0\rangle$ is the vacuum state. From Eqs.~(\ref{hamiltonian}) and (\ref{state}), we obtain the eigenequation of the system
\begin{equation}
-t(\psi_{n+1}+\psi_{n-1})+V_n\psi_n=E\psi_n,\label{ccc}
\end{equation}
from which we can obtain the ground state. One-particle ground states are always extended for periodic systems, which can be localized for the FK model as the on-site potential $V_n$ may lead to aperiodicity.

To study entanglement properties, we first map a state of fermions to that of spins according to Eq.~(\ref{mappp}). 
After the mapping, our state $|\Psi\rangle$ becomes a multi-qubit state, and the pairwise entanglement 
quantified by the concurrence~\cite{Conc} is well-defined. For state (\ref{state}), 
the concurrence for two LFMs $i$ and $j$ is easily found to be~\cite{Conc2}
$
C_{ij}=2|\psi_i\psi_j|.
$
Specifically, when $|\psi_n|=1/\sqrt{N}$, the state becomes the so-called W state~\cite{W} and the concurrence is given by $2/N$. 
In this study, we are more interested in the gross measure of entanglement, the average concurrence~\cite{Lak03}
\begin{equation}
\langle C\rangle=\frac{1}{M}\sum_{i<j}C_{ij}=\frac{1}{M}\left[\left(\sum_{n=1}^N|\psi_n|\right)^2-1\right], \label{aconc}
\end{equation}
which have connections to localization. Here, $M=N(N-1)/2$. We will mainly concentrate the pairwise entanglement of ground states and briefly discuss the bipartite entanglement.

As a first step of numerical calculations, we obtain the configuration for $N$ atoms
by gradient method~\cite{au83}, adopting the periodic  boundary condition.
It is well-known that there exists a critical value $K_c=0.154641$ separating two configurations of atoms. 
The configurations determine the on-site potential $V_n$, and thus ground-state entanglement. 

Figure 1 displays behaviors of the average concurrence (\ref{aconc}) of the ground state of the FK model as a function of $K$.  When the parameter $K$ increases, 
we observe an abrupt decrease of the average concurrence near critical value $K_c$. 
There is a strong interrelation between the electronic properties and the configurations of atoms, and thus this transition of entanglement results from the transition by breaking of analyticity in the configurations of atoms. The entanglement strongly {\it feels} the classical transition. 
For $K<K_c$, the configuration of atoms corresponds to invariance circle. In this case, the concurrences $\langle C\rangle\approx 2/N$, and the ground states are extended. 
For $K>K_c$, the configuration corresponds to cantorus, and ground states are quite different 
from the case of $K<K_c$. In this case, the concurrence tends to disappear and the ground state is localized.	

To see clearly between the entanglement and localization, we show in Fig.~2 the average concurrence against 
the participation ratio.  The participation ratio characterizing the degree of localization is defined by 
\begin{equation}
p=\frac{1}{N\sum_{n=1}^N|\psi_n|^4}.\label{ppp}
\end{equation}
We see that the average concurrence increases with the increase of the participation ratio, 
illustrating that the more localized ground states are, the less the average pairwise entanglement. 

\begin{figure}
\includegraphics[width=0.40\textwidth]{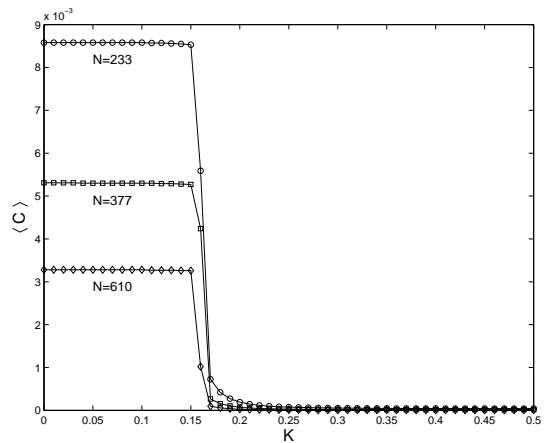}
\caption{Average concurrence against parameter $K$ for FK chains with different lengths. The parameter $\lambda=3$.}
\end{figure}

\begin{figure}
\includegraphics[width=0.40\textwidth]{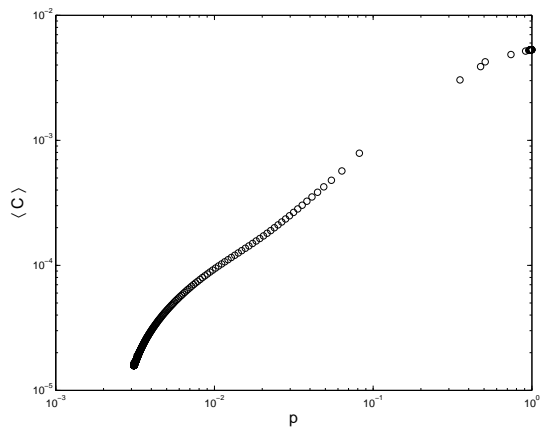}
\caption{Average concurrence against the participation ratio. The parameters $\lambda=3$ and $N=377$.}
\end{figure}

\begin{figure}
\includegraphics[width=0.40\textwidth]{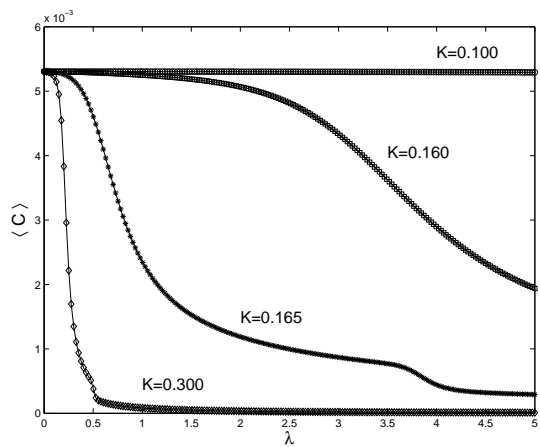}
\caption{Average concurrence against the parameter $\lambda$ for different $K$. The parameter $N=377$.}
\end{figure}

Now, we investigate effects of amplitude $\lambda$ of the on-site potential on ground-state entanglement. The average concurrence 
as a function of $\lambda$ is shown in Fig.~3 for different $K$. When $K=0.1$, the concurrence is nearly unchanged, 
and approximately given by $2/377=0.005305$. As $K<K_c$, atoms are in the configuration before the breaking of analyticity, 
and this configuration is nearly the same as that for $K=0$. Then, the on-site potential $V_n=\lambda\cos(2\pi\sigma x_n^0)$ are approximately of no difference, and the second term of Eq.~(\ref{hamiltonian}) contribute a constant to the Hamiltonian for one-particle states. The concurrence of the ground state is then only determined by the first term of Eq.~(\ref{hamiltonian}), and thus independent of $\lambda$. When $K=0.3$, the concurrence reduces quickly to zero as $\lambda$ increases. In this case, the atoms are in the configuration after the breaking of analyticity, and the second term of Eq.~(\ref{hamiltonian}) 
no longer commute with the first one and has significant effects on the ground-state properties.
For larger $\lambda$, the second term dominate over the first one, and thus reduce the entanglement. 
It is evident that the concurrence goes to zero as $\lambda\rightarrow\infty$. The curves for $K=0.16$ and $0.165$ 
displays the intermediate behaviors. For $K=0.3$, we observe a critical value $\lambda_c$, after which
the entanglement tends to zero. The critical value is dependent on the parameter $K$.

So far, we have studied ground-state entanglement properties in the FK model. 
Next, we investigate dynamics of entanglement with the initial state being $|1\rangle=c_1^\dagger |0\rangle$. Thus, there is no initial entanglement. The time evolution is described by a time-dependent equation ($\hbar=1$)
\begin{equation}
i\frac{d\psi_n}{dt}=-\psi_{n+1}-\psi_{n-1}+\lambda\cos(2\pi\sigma x_n^0)\psi_n,
\end{equation}
which can be integrated numerically by various methods such as the fourth-order Runge-Kutta method.

Figure 4 displays dynamical behaviors of the average concurrence for different $K$ and $\lambda$. From Fig.~1 (a), (b) and (c) for $K=0.1$,  
we observe that the concurrence first linearly increases with time, and finally reaches a plateau, oscillating irregularly around a 
steady value. The dynamical behaviors only differ slightly for different $\lambda$. The reason is similar to that discussed above, 
namely, for this case the term $\lambda\cos(2\pi\sigma x_n^0)$ contributes a constant  to the Hamiltonian and does not alter 
the system dynamics. In contrast, for $K=0.3$, as seen from Fig.~1 (d), the increase of $\lambda$ suppresses the entanglement generation. When $\lambda$ becomes larger, 
the electronic states becomes more localized, and the entanglement diminishes. 
The slight dependence of $\lambda$ for $K=0.1$ and strong dependence of $\lambda$ for $K=0.3$ corresponds to extended and localized ground states, respectively.

\begin{figure}
\includegraphics[width=0.40\textwidth]{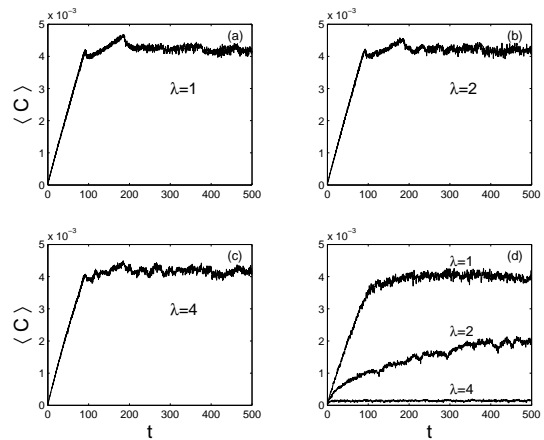}
\caption{Time evolution of the average concurrence. The parameter $N=377$. (a), (b) and (c) correspond to  $\lambda=1$, $\lambda=2$, and $\lambda=4$, respectively; $K=0.1$. (d) corresponds to $K=0.3$ for different $\lambda$. The parameter $N=233$.}
\end{figure}

From the above results on ground-state and dynamical behaviors of entanglement, 
we see that the more localized a state is, the less the entanglement. 
Now, we build a direct connection between the concurrence, quantifying the pairwise entanglement, 
and the participation ratio, characterizing the degree of localization. For one-particle state $|\Psi\rangle$ (\ref{state}), 
the concurrence between LFMs $i$ and $j$ $C_{ij}=2|\psi_i\psi_j|$. We make an average of square of concurrence,
rather than the concurrence,
\begin{align}
\langle C^2\rangle=&\frac{1}{M}\sum_{i<j}(C_{ij})^2=\frac{4}{M}\sum_{i<j}|\psi_i|^2|\psi_j|^2\nonumber\\
=&\frac{2}{M}\left(1-\sum_{i=1}^N|\psi_i|^4\right)\nonumber\\
=&\frac{4}{N(N-1)}\left(1-\frac{1}{Np}\right).\label{relation}
\end{align}
In deriving of the above equation, we have used Eq.~(\ref{ppp}) and the identity 
$
2\sum_{i<j}|\psi_i|^2|\psi_j|^2=1-\sum_{i=1}^N|\psi_i|^4,
$ 
which results form the normalization condition $\sum_{i=1}^N|\psi_i|^2=1$.
Thus, the average of the square of concurrence can be written as a simple function of the participation ratio, and this relation 
build a direct connection between pairwise entanglement and localization. It is evident that the larger $p$ is, the larger the concurrence.
For the two extreme cases, $p=1/N$ and $p=1$, the concurrence $C$ becomes 0 and $4/N^2$, respectively, as we expected. Note that the relation is applicable to arbitrary one-particle states, irrespective of model Hamiltonians.

We have discussed the pairwise entanglement. For other type of entanglement, such as the bipartite pure-state entanglement quantified by the linear entropy, we also find similar relations as Eq.~(\ref{relation}) between the linear entropy and participation ratio~\cite{HKBU}. The connections between entanglement and localization are not restricted to electronic system, and can be applied to other systems such as spin systems with one-magnon excitations. The investigation of multipartite entanglement other than pairwise and bipartite entanglement is more interesting, but at the same time more difficult and complicated. 

In conclusion, we have studied ground-state and dynamical pairwise entanglement of two LFMs 
in the one-dimensional FK model, and found that the entanglement exhibits distinct behaviors 
for the cases of $K<K_c$ and $K>K_c$. This is a consequence of the transition by breaking of analyticity. For $K<K_c$, the 
ground state is extended and more entangled; while for $K>K_c$ the ground state is localized and less or not entangled. 
The amplitude $\lambda$ of the on-site potential have slight effects on entanglement when $K<K_c$, while has 
significant effects when $K>K_c$. It is interesting to note that entanglement is closely 
connected to localization. It becomes a general feature that the more extended the electron is, the more entangled the electronic state. 

Our results support the idea that quantum information theory offers a powerful approach to the study of nonlinear complex system. 
At the transition point, the concurrence is strongly affected, just as behaviors of the concurrence in the 
quantum phase transition point~\cite{qpt1,qpt2}. It would be more attractive to study 
entanglement behaviors in other nonlinear complex systems such as two-dimensional quasicrystals and disorder systems.

\acknowledgments We acknowledge valuable discussions with L. Yang, X.W. Hou and Z.G. Zheng. 
This work was supported by the grants from the Hong Kong Research Grants Council (RGC) and the Hong Kong
Baptist University Faculty Research Grant (FRG). X. Wang has been supported by an Australian 
Research Council Large Grant and Macquarie University Research Fellowship.

\end{document}